\documentclass[12pt]{article}
\usepackage{graphicx}

\newcommand\pubnumber{SNSN-323-63}
\newcommand\pubdate{\today}

\def\institute{Faculty of Mathematics, Physics and Informatics\\
Comenius University, SK-84248 Bratislava, SLOVAKIA}
\def\support{\footnote{Copyright 2019 CERN for the benefit of the ATLAS and CMS Collaborations. Reproduction of this article or parts of it is allowed as specified in the CC-BY-4.0 license.}}
\usepackage{wrapfig}

\def\Title#1{\begin{center} {\Large #1 } \end{center}}
\def\Author#1{\begin{center}{ \sc #1} \end{center}}
\def\Address#1{\begin{center}{ \it #1} \end{center}}

\newcommand\pubblock{\rightline{\begin{tabular}{l} \pubnumber\\
         \pubdate  \end{tabular}}}
\newenvironment{Abstract}{\begin{quotation}  }{\end{quotation}}
\newenvironment{Presented}{\begin{quotation} \begin{center} 
             PRESENTED AT\end{center}\bigskip 
      \begin{center}\begin{large}}{\end{large}\end{center} \end{quotation}}
\def\Acknowledgements{\bigskip  \bigskip \begin{center} \begin{large}
             \bf ACKNOWLEDGEMENTS \end{large}\end{center}}

\def\beq{\begin{equation}}
\def\eeq#1{\label{#1}\end{equation}}
\def\eeqn{\end{equation}}

\def\beqa{\begin{eqnarray}}
\def\eeqa#1{\label{#1}\end{eqnarray}}
\def\eeqan{\end{eqnarray}}

\let\bar=\overbar

\def\Dslash{\not{\hbox{\kern-4pt $D$}}}
\def\dslash{\not{\hbox{\kern-2pt $\del$}}}

\def\msb{{\bar{\ssstyle M \kern -1pt S}}}

\begin{document}
\begin{titlepage}
\pubblock

\vfill
\Title{Top-quark mass at ATLAS and CMS}
\vfill
\Author{ Stanislav Tokar On behalf of the ATLAS and CMS Collaborations.\support}
\Address{\institute}
\vfill
\begin{Abstract}
The top-quark mass measurements carried out by the LHC experiments, ATLAS and CMS, are summarized. Results of different approaches to the top-quark mass reconstruction are presented. Masses from different measurements are in good agreement within uncertainties. Precision of the measurements with the directly measured top-quark mass is now better than 0.5\%. Progress in determination of the top-quark pole mass is reported and the relation between the directly measured top-quark mass and top-quark pole mass is discussed. 
\end{Abstract}
\vfill
\begin{Presented}
$11^\mathrm{th}$ International Workshop on Top Quark Physics\\
Bad Neuenahr, Germany, September 16--21, 2018
\end{Presented}
\vfill
\end{titlepage}
\def\thefootnote{\fnsymbol{footnote}}
\setcounter{footnote}{0}

\vspace{-4mm}
\section{Introduction}
\vspace{-2mm}
The top quark is the heaviest particle of the Standard Model (SM) and its mass, $m_\mathrm{t}$, is an important SM parameter. Precise
measurements of $m_\mathrm{t}$ provide critical inputs to fits of global electroweak parameters \cite{Baak2014} that test the internal
consistency of the SM. In addition, the value of $m_\mathrm{t}$ strongly affects the quartic coupling of the SM Higgs potential, which has cosmological implications \cite{PRep179_273}. 
The Large Hadron Collider (LHC) experiments, ATLAS \cite{atl_det} and CMS \cite{cms_det}, collect data on top quarks since 2010. After stopping Tevatron in 2011, the LHC is the only machine capable of producing top quarks. 
The abundance of the top-quark data collected by the LHC experiments in proton-proton ($pp$) collisions at a centre-of-mass energy of $\sqrt{s}$ = 7, 8 and 13 TeV, has enabled to measure the top-quark mass in different ways with high precision.

An important issue concerning top-quark mass measurements is what kind of mass is measured in the experiments. The reason is that theoretical predictions usually use the top-quark pole mass which exhibits an intrinsic ambiguity of the order of $\Lambda_\mathrm{QCD}$ due to the fact that top quark is not an asymptotically free particle.
 On the other hand, on the experimental side there are limited possibilities to separate the top-quark decay products from the rest of the interaction, e.g. in the hadronization of the $b$-quark from a top-quark decay at least one quark coming from the rest of the interaction is captured. A discussion on this subject can be found in Refs. \cite{hoang14, 4loops, nason17}. 

\vspace{-3mm}
\section{Top-quark mass reconstruction}
\noindent Presently, the top-quark mass is inferred in two basic ways. The first approach employs  kinematic observables sensitive to the top-quark mass -- it can be e.g. the invariant mass $m_t$ reconstructed from the top-quark decay products (kinematic approach).
In this case different techniques \cite{kondo88, cdf_tem2006, PRL98}
are used to extract the top-quark mass (kinematic top-quark mass).   The second approach employs the relation of the top-quark pole mass and the top-quark pair ($t\bar{t}$) production cross section.
The more precise results are obtained within the former (kinematic) approach, but in this approach the renormalization scheme is not well defined and thus it is not clear what is the relation between the measured mass and the top-quark pole mass. 
For this reason, a lot of effort is devoted to the latter approach, where the mass is inferred from the measured $t\bar{t}$ cross section. Results of both these approaches will be reported.

\vspace{-3mm}
\subsection{Top-quark mass reconstructed at LHC}
\vspace{-2mm}
The LHC experiments, ATLAS and CMS, have measured the top-quark mass at all up-to-now available centre-of-mass energies $\sqrt{s}=$ 7, 8 and 13 TeV  using different approaches. This enabled  a large progress in precision of the measured top-quark  mass.  

{\bf Top-quark mass at ATLAS.} 
ATLAS has measured the top-quark mass at $\sqrt{s}=$7 and 8 TeV.  The measurements performed at 8 TeV in lepton+jets ($\ell$+jets), dilepton ($\ell\ell$) and all-jets channels are reported. The measurements, carried out 
\begin{wrapfigure}{r}{0.45\textwidth}
\vspace{-3mm}
\centering
\begin{tabular}{c}
\includegraphics[width=0.40\textwidth]{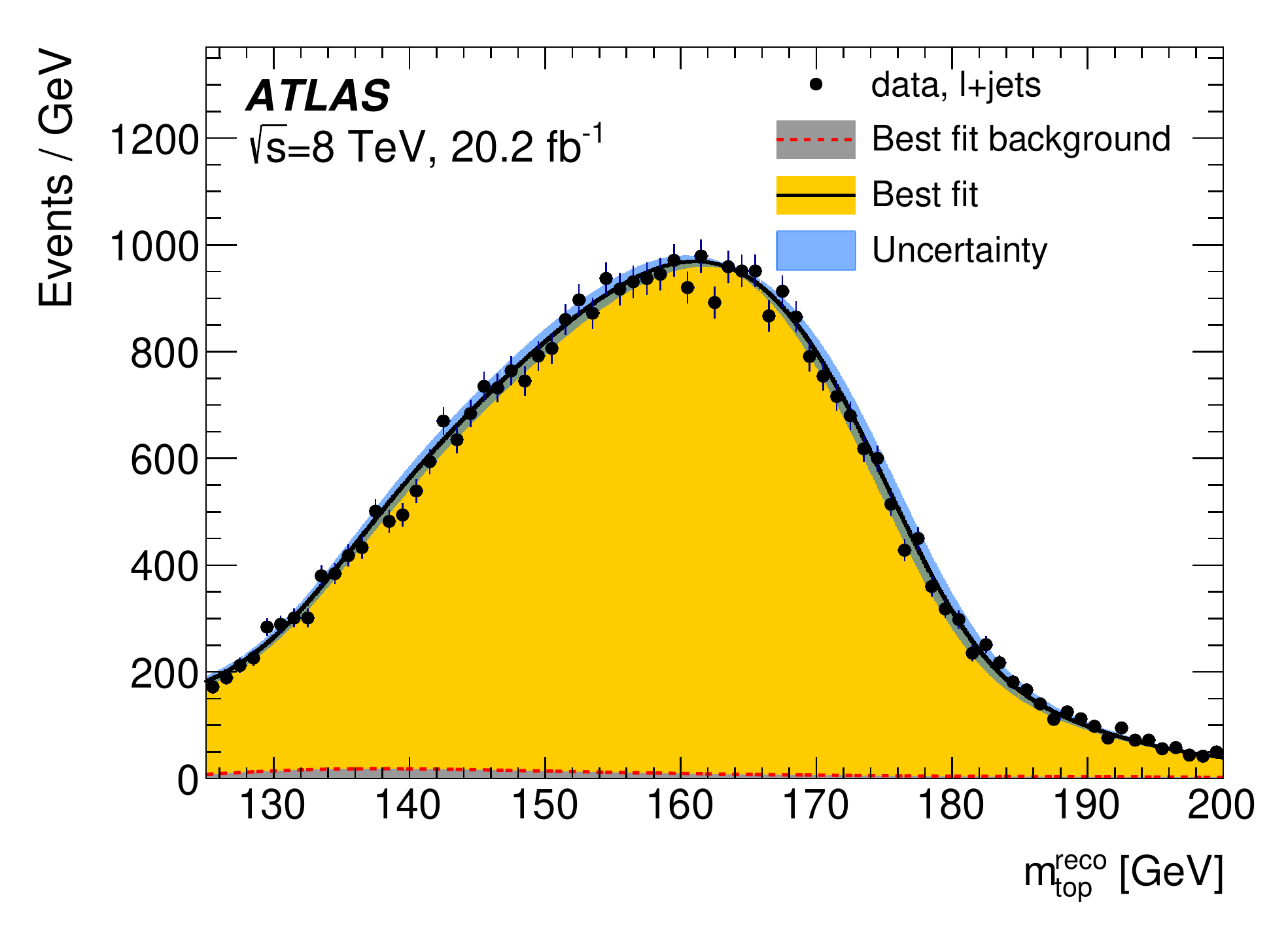}
\end{tabular}
\vspace{-4mm}
\caption{
\small{The distribution of $m\mathrm{_{top}^{reco}}$ and the best fit probability density function for the
background alone and for the signal+background hypothesis\cite{atl_conf-2017-071}.}}
\label{fig:atl_LJ_fig6a}
\end{wrapfigure}
\noindent in the $\ell$+jets  channel, employed a fit to 3D templates derived from MC at different $m_{t}$  \cite{atl_conf-2017-071}. 
The signal and background templates of three observables, $m\mathrm{_{top}^{reco}}$, $m_W^\mathrm{reco}$, and $R_{bq}^\mathrm{reco}$ are used in an unbinned fit to the selected data events. The observable $m\mathrm{_{top}^{reco}}$ is the reconstructed invariant mass of the top-quark decay products, $m_W^\mathrm{reco}$ is the invariant mass of the hadronically decaying $W$ boson, and $R_{bq}^\mathrm{reco}$ is a ratio of the transverse momentum of the $b$-tagged jet to the average transverse
momentum of the two jets of the hadronic $W$ boson decay. 
The outputs of the likelihood fit are the reconstructed top-quark mass, $m_t$, the jet energy scale factor, JES, and a relative $b$-jet energy scale factor, $b$JSF. As a representative fit result, the fitted data distribution is shown in Fig. \ref{fig:atl_LJ_fig6a} showing an $m\mathrm{_{top}^{reco}}$ distribution. 
 The extracted value of  $m_t$ is
\vspace{-1.5mm}
\begin{tabbing}
\indent $m_t^{\ell\mathrm{+jets}}$ = 172.08 $\pm $ 0.39 (stat) $\pm $0.82 (syst) GeV.
\end{tabbing}
\vspace{-1.5mm}
In the dilepton case, a 1D template method was based on the sensitive observable $m_{\ell b}^\mathrm{reco}$, i.e. the lepton-$b$-jet invariant mass  \cite{atl_PLB761_2016}. An unbinned likelihood fit to the data was based on $m_{\ell b}^\mathrm{reco}$ signal and background templates and the outputs of the fit were the top-quark mass, $m_t^{\ell\ell}$, and the background fraction, $f_\mathrm{bkg}$. The extracted $m_t^{\ell\ell}$ is 
\vspace{-2mm}
\begin{tabbing}
\indent $m_t^{\ell\ell}$ = 172.99 $\pm $ 0.41 (stat) $\pm $ 0.74 (syst) GeV.
\end{tabbing}
\vspace{-2mm}
The combination of the ATLAS top-quark measurements in the dilepton and $\ell$ + jets channels at  $\sqrt{s}$ = 7 and 8 TeV using the BLUE technique \cite{BLUE_2014} gives:
\vspace{-2mm}
\begin{tabbing}
\indent  $m_t^\mathrm{comb}$ = 172.51 $\pm $ 0.27 (stat) $\pm $ 0.42 (syst) GeV.
\end{tabbing}
\vspace{-2mm}
The relative precision of the combined result is 0.29\% 
-- see details in Ref.\cite{atl_conf-2017-071}.

\noindent The template method was also used for the measurement of the top-quark mass in the all-jets channel at $\sqrt{s}$ = 8 TeV \cite{atl_all-jets_8TeV_jhep2017}. The sensitive observable is $R_{3/2} = m_\mathrm{jjj}/m_\mathrm{jj}$, where $m_\mathrm{jjj}$ ($m_\mathrm{jj}$) is the three- (two-) jet invariant mass corresponding to the top-quark  ($W$ boson) decay products. $R_{3/2}$ is used 
to reduce the systematic effects common to the reconstructed top-quark and $W$ boson masses. 
The large multijet background was estimated using a data driven technique. A template fit to $R_{3/2}$ with a binned minimum-$\chi^2$ was applied providing as output the top-quark mass $m_t^\mathrm{had}$ and background fraction $F_\mathrm{bkg}$. The extracted $m_t^\mathrm{had}$ reads
\vspace{-2mm}
\begin{tabbing}
\indent $m_t^{\mathrm{had}}$ = 173.72 $\pm $ 0.55 (stat) $\pm $ 1.01 (syst) GeV.
\end{tabbing}
\vspace{-2mm}
The precision is around 40\% better than the previous one at $\sqrt{s}$ = 7 TeV \cite{atl_EPJC75_2015}. 

{\bf Top-quark mass at CMS.}  The CMS experiment measured the top-quark mass at $\sqrt{s}$~=~7,~8~and 13 TeV. The measurements at $\sqrt{s}$~=~13 TeV performed in the $\ell$+jets
\begin{wrapfigure}{r}{0.44\textwidth}
\vspace{-3mm}
\centering
\begin{tabular}{c}
\includegraphics[width=0.40\textwidth]{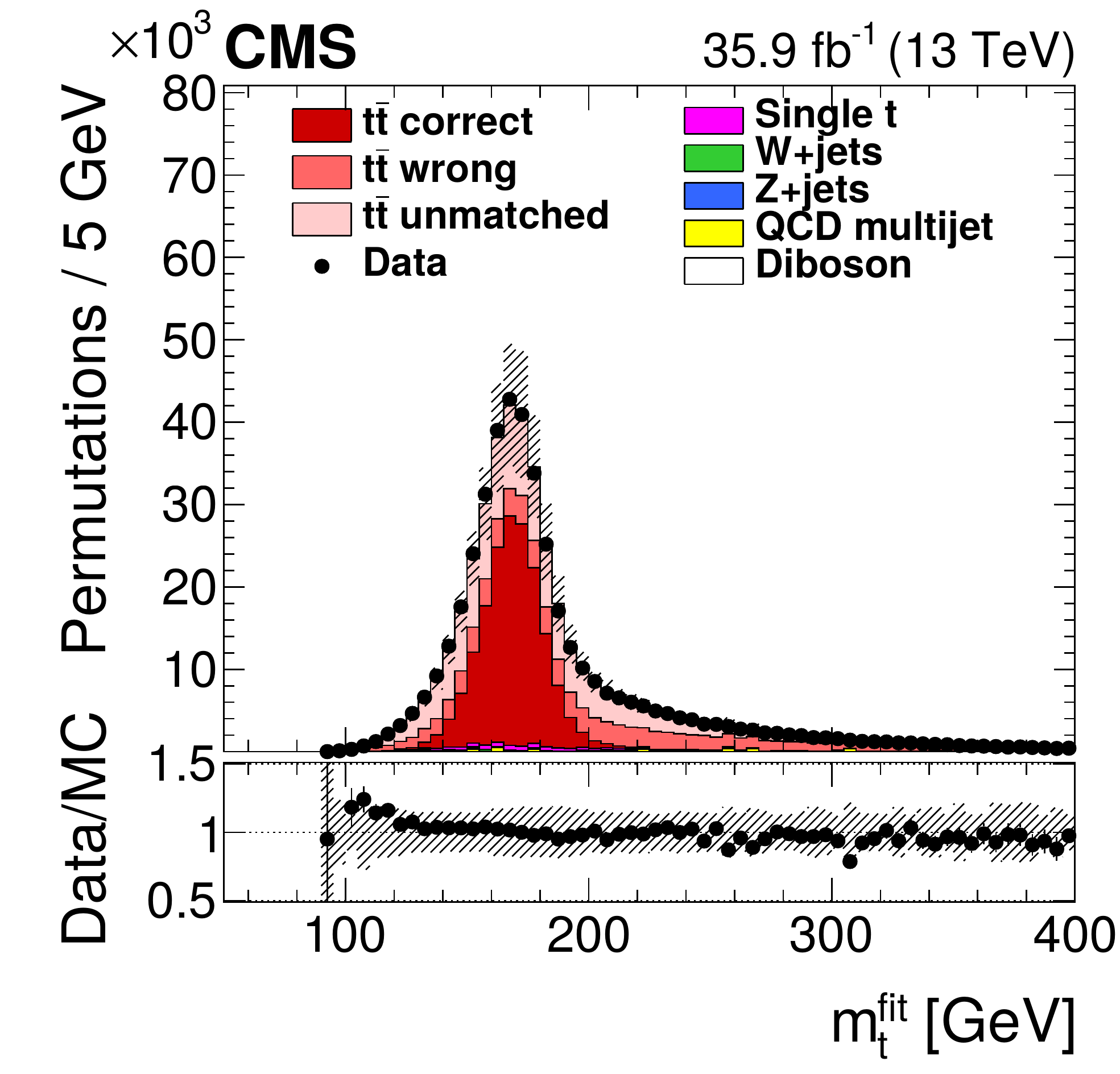}
\end{tabular}
\vspace{-5mm}
\caption{
\small{Fitted top-quark masses $m_t^\mathrm{fit}$ after the goodness-of-fit selection and the weighting by $P_\mathrm{gof}$ (see text) \cite{cms_LJ13TeV}. }
}
\label{fig:cms_LJ13TeV_fig2}
\vspace{-4mm}
\end{wrapfigure}
\noindent   and all-jets channels, and also the measurement in the dilepton channel at $\sqrt{s}$~=~8~TeV as well as the combined 7 and 8 TeV result, are presented.
The $\ell$+jets measurement carried out using the data sample of 35.9~fb$^{-1}$ employs an ideogram technique \cite{cms_LJ13TeV} which 
is based on a joint maximum likelihood fit to data. The fit output is the top-quark mass, $m_t$, and (optionally) the jet energy scale factor, JSF. The likelihood fit is based on an event likelihood created using $m_t^\mathrm{fit}$  and  $m_W^\mathrm{reco}$ templates obtained from a simulation for different $m_t$ and JSF. 
The sensitive observables are the masses $m_t^\mathrm{fit}$  and  $m_W^\mathrm{reco}$ corresponding to top quark and $W$ boson, respectively, which are estimated by a kinematic fit for each event and different parton-jet assignments. In addition, to each jet permutation a probability $P_\mathrm{gof}$ is assigned. Three approaches are used to reconstruct the top-quark mass: {\it 2D approach} with a simultaneous fit to 

\begin{wrapfigure}{r}{0.44\textwidth}
\vspace{-3mm}
\centering
\begin{tabular}{c}
\includegraphics[width=0.42\textwidth]{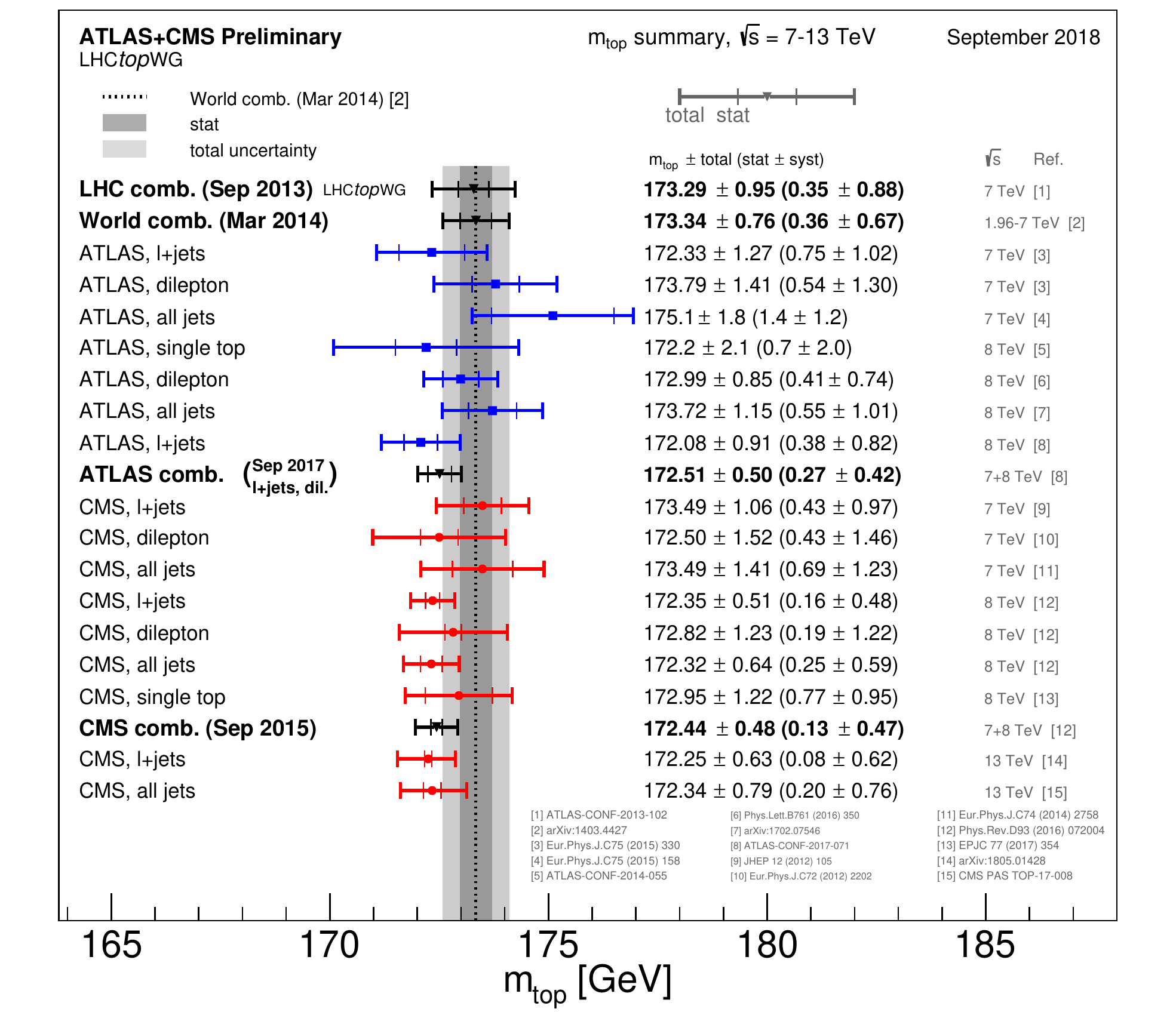}
\end{tabular}
\vspace{-7mm}
\caption{
{\small Summary of the ATLAS and CMS direct $m_t$ measurements \cite{atl_top-web}.} 
}
\label{fig:topMass_summary}
\vspace{-10mm}
\end{wrapfigure}
\noindent  $m_t$ and JSF,
{\it 1D approach} with a fit only to $m_t$ (JSF = 1) and {\it hybrid approach} with a prior knowledge about the JES is used but a Gaussian constrain is applied to its value to account for the JES uncertainty.
Fig. \ref{fig:cms_LJ13TeV_fig2} shows the final distributions, after the $P_\mathrm{gof}$ selection, of the invariant mass, $m_t^\mathrm{fit}$, of the top-quark candidates for all selected  
jet permutations (see Ref. \cite{cms_LJ13TeV}). The most precise result is obtained in the hybrid approach (precision of 0.37\% ):
\vspace{-5mm}
\begin{tabbing}
{\small \noindent $m_t^{\ell+\mathrm{jets}}$ = 172.25 $\pm $ 0.08 (stat) $\pm $ 0.62 (syst) GeV.}
\end{tabbing}
\vspace{-1mm}

\noindent \noindent The same ideogram technique 
was used also for the  measurement carried out in the all-jets 
channel \cite{cms_allJets-13TeV}. The most precise mass is again obtained with the hybrid approach:
\vspace{-5mm}
\begin{tabbing}
{\small \indent $m_t^{\mathrm{all}-\mathrm{jets}}$ = 172.35 $\pm $ 0.25 (stat) $\pm $ 0.59 (syst) GeV.}
\end{tabbing}
\vspace{-1mm}

\noindent In the $\ell\ell$ case an analytical matrix weighting technique (AMWT) - see details in Ref. \cite{cms_DiL-8TeV_prd93} - was employed. The likelihood fit to the data gives 
\vspace{-2mm}
\begin{tabbing}
\indent {\small $m_t^{\ell\ell}$  = 172.82 $\pm $ 0.19 (stat) $\pm $ 1.22 (syst) GeV.}
\end{tabbing}
\vspace{-2mm}
\noindent The CMS collaboration combined its top-quark mass measurements at 7 and 8 TeV taking into account correlations between them. The obtained mass is
\vspace{-2mm}
\begin{tabbing}
\indent {\small $m_t^\mathrm{comb}$  =  172.44 $\pm $ 0.13 (stat) $\pm $ 0.47 (syst)  GeV.}
\end{tabbing}
\vspace{-2mm}
The relative precision of the combination is 0.28\%.       
 A summary of the direct ATLAS and CMS top-quark mass measurements is given in Fig. \ref{fig:topMass_summary}. 
                                       
\vspace{-4mm}
\section{Top-quark pole mass} 
\vspace{-3mm}
\noindent The dependence of the $t\bar{t}$ cross section 
 on the top-quark pole mass, $m_t^\mathrm{pole}$, is used to infer this mass. 
From the theoretical point of view the $m_t^\mathrm{pole}$, unlike the directly reconstructed top-quark mass, is a well-defined parameter and a big progress in the $t\bar{t}$ cross section calculations 
\cite{Czakon_PRL13}, makes the idea of its determination  very attractive. 

{\bf Top-quark pole mass at ATLAS.} For the extraction of $m_t^\mathrm{pole}$ the ATLAS
experiment used  differential $t\bar{t}$ cross section measurements at 8 TeV performed in dilepton $e/\mu $ channel with  electron and muon as decay products \cite{atl_topPoleM}. The pole mass is extracted from eight normalized differential $t\bar{t}$ cross sections by comparison of the normalized measured differential cross sections and theoretical predictions parametrized as  functions of $m_t^\mathrm{pole}$ using a minimal $\chi ^2$ approach. 
The extracted pole mass reads
\begin{center}
\vspace{-2mm}
 $m_t^\mathrm{pole}$  = 173.2 $\pm $ 0.9 (stat) + 0.8(syst) +1.2 (theo) GeV.
\vspace{-2mm}
\end{center}
The main systematic uncertainties come from $t\bar{t}$ modelling, QCD scale, PDFs, lepton efficiencies and jets/$b$-tagging. 

{\bf Top-quark pole mass at CMS.} 
The CMS collaboration has extracted the top-quark pole mass, $m_t^\mathrm{pole}$, from the inclusive $t\bar{t}$ cross section ($\sigma_{t\bar{t}}$) 
\noindent measurements at 7 and 8 TeV performing the measurements  in the dilepton $e/\mu $ channel 
\cite{cms_topPoleM}.  
The predicted NNLO+NNLL $t\bar{t}$ production cross sections, employing NNPDF3.0 and $\alpha_{S}$ = 0.118 $\pm $ 0.001, were compared to the measured $\sigma_{t\bar{t}}$ at $\sqrt{s}$ = 7 and 8 TeV as a function of $m_t^\mathrm{pole}$. 
The pole mass was extracted at each value of $\sqrt{s}$
and the combined mass reads
\begin{center}
\vspace{-2mm}
 $m_t^\mathrm{pole}$  = 173.8 $_{-1.7}^{+1.8} $  GeV.
\vspace{-2mm}
\end{center}
The extracted masses using the CT14 and the MMHT2014 PDF sets give, within uncertainties, compatible results. The main systematic uncertainties come from $t\bar{t}$ modelling and QCD scale, from PDFs, lepton efficiencies, jets and $b$-tagging. 

{\bf Top-quark pole mass from inclusive $\mathbf{t\bar{t}}$+1 jet cross section.} 
The $t\bar{t}$+1-jet cross section ($\sigma_{t\bar{t}+\mathrm{jet}}$) exhibits an interesting (enhanced with respect to $\sigma_{t\bar{t}}$) dependence on  $m_t^\mathrm{pole}$ \cite{JHEP10-2015}. The top-quark pole mass can be extracted from the normalized differential distribution:
\vspace{-2mm}
\begin{equation}
\centering
R\left(m_t^\mathrm{pole},\rho _s\right)=\frac{1}{\sigma_{t\bar{t}+\mathrm{jet}}}\frac{d\sigma_{t\bar{t}+\mathrm{jet}}}  {d\rho_s}\left(m_t^\mathrm{pole},\rho _s\right),
\label{poleM_rho}
\end{equation}
\vspace{-1mm}
\noindent where $s_{t\bar{t}j}$ is the $t\bar{t}j$ invariant mass and $m_0$ is an arbitrary mass (usually $m_0 \sim $ 170~GeV). 
Then a template technique is used to extract $m_t^\mathrm{pole}$.  

The ATLAS collaboration measured the top-quark pole mass at $\sqrt{s} = $ 7 TeV with the integrated luminosity of 4.6 fb$^{-1}$ \cite{atl_poleM-2015}. The extracted mass value is:
\begin{center}
\vspace{-2mm}
 $m_t^\mathrm{pole}$  = 173.7 $\pm $ 1.5 (stat) + 1.4(syst) + $_{-0.5}^{+1.0} $  (theo) GeV.
\vspace{-2mm}
\end{center}
The main background comes from Single top, W/Z+jet and fake leptons. The main sources of systematics are QCD scale 
($\mu_\mathrm{R}$,  $\mu_\mathrm{F}$) variation, JES, ISR/FSR and PDFs. 

CMS performed a similar analysis based on the observable $\rho_s$ in the dilepton channel at $\sqrt{s} = $ 8 TeV with the data sample of 19.7 fb$^{-1}$ and obtained the mass \cite{cms_top-13-006}:
\begin{center}
\vspace{-2mm}
 $m_t^\mathrm{pole}$  = 173.7 $\pm $ 1.5 (stat) + $_{-3.1}^{+2.5}$(syst) + $_{-1.6}^{+3.8} $ (theo) GeV.
\vspace{-2mm}
\end{center}
Also in this case the main systematic uncertainties come from $\mu_\mathrm{R}$ and $\mu_\mathrm{F}$ scale variation, jet-parton matching, hadronization and color reconnection.

\vspace{-4mm}
\section{Summary}
In the ATLAS and CMS experiments, the top-quark mass is investigated with a variety of approaches giving compatible results among them and now its uncertainty is deeply below 1 GeV (kinematic top-quark mass) and approaches to  $\Lambda_\mathrm{QCD}$. A big effort on both the experimental  and theoretical side continues move us toward  a better understanding of this fundamental SM parameter.  
 
\vspace{-6mm}
\Acknowledgements
I am grateful to the top quark and top-quark mass groups of the ATLAS and CMS collaborations for their comments. 

\end{document}